\documentstyle[12pt,epsfig]{article}
\def\beq{\begin{equation}}
\def\eeq{\end{equation}}
\def\bea{\begin{eqnarray}}
\def\eea{\end{eqnarray}}

\begin{document}
\pagestyle{empty}
\begin{flushright}
{DSF 44/04}\\
\end{flushright}
\vspace*{5mm}
\begin{center}
{\bf SPECTRUM OF THE Y=2 PENTAQUARKS}
\\
\vspace*{0.5cm}
{\bf F. Buccella, D. Falcone and F. Tramontano} \\
Dipartimento di Scienze Fisiche, Universit\`{a} di Napoli ``Federico II",
Via Cintia, I-80126 Napoli, Italy \\
INFN, Sezione di Napoli

\vspace{0.1cm}
\vspace*{1.5cm}

{\bf ABSTRACT} \\ \end{center}

\vspace*{2mm}

\noindent
By assuming a mass formula for the spectrum of the $Y=2$ pentaquarks,
where the chromo-magnetic interaction plays a main role, and identifying
the lightest state with the $\Theta^+(1540)$, we predict a spectrum in
good agreement with the few $I=0$ and $I=1$ candidates proposed in the past. 

\vspace*{1.3cm}



\begin{flushleft}
\end{flushleft}
\newpage
\setcounter{page}{2}
\pagestyle{plain}

\section{Introduction}

The quark model is a successful theory, which is widely used to classify
hadrons and calculate their masses and decays. According to the simple
quark model, mesons are formed by a quark and an antiquark of the opposite
colour, and baryons by three quarks in a colour singlet. However, besides
these states, QCD does not exclude the existence of multiquark 
(tetraquark, pentaquark,...) states, which may transform as higher $SU(3)_F$
representations.\\
In fact, recently, the discovery of a narrow ($Y=2, I=0$) KN resonance $\Theta^+$ 
at 1540 MeV \cite{N}, just the value predicted \cite{DPP} in the Skyrme model \cite{S}, 
has motivated the search \cite{SR} \cite{JW} \cite{JM} \cite{BGS} \cite{CCKN1} 
\cite{BS} \cite{CD} \cite{D} \cite{HS3} to understand it as a $uudd\bar{s}$ state.
A study of pentaquarks has been performed many years ago 
\cite{HS}, at the time when bubble chambers, the best device to detect 
these particles, were working. To get narrow widths, one considered 
\cite{HS} P-wave states with the constituents separated by the orbital
centrifugal barrier in two clusters such that the $\bar{q}$ could not form
a colour singlet with a quark of its cluster, as 
$(qq\bar{q})^{\bar{6}}(qq)^6$ or $(qqq)^8 (q \bar{q})^8$.\
The contribution of the chromo-magnetic force, which accounts for the 
mass splittings within the $SU(6)_{FS}$ multiplets \cite{DGG}, was 
supposed to be the sum of the ones coming from the two clusters, since the
centrifugal barrier suppresses the contact interaction.\\
In the following section we shall propose a mass formula for the positive
parity pentaquarks built with $4q,L=1$ and a $\bar{q}$ in 
S-wave with respect to them or with $4q,L=0$ and a
$\bar{q}$ in P-wave with respect to them. We shall extend our consideration
to the negative parity states built with $4q$ and a
$\bar{q}$ in S-wave.\\

\section{A mass formula for pentaquarks}

In \cite {JW} the authors consider two $2q$ clusters transforming as the 
$(\bar{3},1,\bar{3})$ of $SU(3)_C \times SU(2)_S \times SU(3)_F$ in 
P-wave, which implies the most negative contribution from the 
chromomagnetic interaction for the $4q,L=1$ states transforming as a 
$(3,1,\bar{6})$ representation.
In fact the contribution for the interaction of two quarks is given by:

\bea 
\Delta m_{qq} 
= -C_{qq} \left[ C_6(2q) - \frac{1}{2} C_3(2q) - \frac{1}{3} C_2(2q) - 4 \right]
\label{equation1}
\eea
where $C_6(2q)$, $C_3(2q)$ and $C_2(2q)$ are the $SU(6)_{CS}$, $SU(3)_C$ and $SU(2)_S$
Casimir operators, respectively. The consequences of eq.(1) are reported in Table 1

\begin{table}
\begin{center}
\begin{tabular}{|c|c|}
\hline
& \\[-9pt]
$SU(3)_C \times SU(2)_S$   &   $\frac{\Delta m_{qq}}{C_{qq}}$ \\[3pt]
& \\[-9pt]
\hline
& \\[-9pt]
$(\bar{3},1)$              &    $-2$\\[3pt]
$(6,3)$                    &    $-\frac{1}{3}$ \\[3pt]
$(\bar{3},3)$              &    $+\frac{2}{3}$ \\[3pt]
$(6,1)$                    &    $+1$\\[3pt]
\hline
\end{tabular}
\caption{Chromomagnetic splittings for $2q$ states.}
\end{center}
\end{table}

The contribution  of the chromo-magnetic interaction of the $4q$ with the
$\bar{q}$, the $\bar{s}$ for the $Y=2$ states, depends on their relative position.
In the approach developed in \cite{HS} ($qq \bar{q}$)($qq$) clusters 
where considered.
Starting with $4q,L=1$ states, one has the same overlap of the $\bar{q}$ 
with the two $qq$ clusters, which implies: 

\bea
\Delta m_{4q,\bar{q}} & = &  C_{4q,\bar{q}}
\left[ C_6(p)-C_6(t)-\frac{1}{3}C_2(p)+\frac{1}{3}
C_2(t)-\frac{4}{3} \right] 
\label{equation2}
\eea
where $C_6(p)$ and $C_6(t)$ are  the Casimir of $SU(6)_{CS}$, 
$p$ and $t$ are the representations for pentaquark and $4q$ states,
respectively, and $C_2(p)$ and $C_2(t)$ are the Casimir of $SU(2)_S$.\\
To account for the low value of the mass of $\Theta^+$, despite the expected
contribution of the kinetic energy associated to the orbital angular momentum, 
we may assume that the overlap of the $\bar{s}$ with the quarks is the same as
in $K^{(*)}$ particles; we take:
\bea
C_{4q,\bar{q}} = \frac{3}{16}( m_{K^*} - m_K)
\label{equation3}
\eea
To compute the r.h.s. of eq.(2), one needs to know the $SU(6)_{CS}$ transformation
properties of each pair of two $q$ clusters in P-wave, which, if
the two clusters have the same transformation properties, depends on the 
$SU(3)_F$ behaviour of the $4q$'s. This information is described in Table 2
together with the contribution of the chromo-magnetic interaction to the
mass splitting.

\begin{table}
\begin{center}
\begin{tabular}{|c|c|c|c|}
\hline
& & & \\[-11pt]
$2q \times 2q$ & & & \\
$SU(3)_C \times SU(3)_S$ & $ SU(6)_{CS}$ & $SU(3)_F \times SU(2)_S$ & $\frac{\Delta m}{C_{qq}}$ \\[3pt]
& & & \\[-11pt]
\hline
& & & \\[-9pt]
$[(\bar{3},1) \times (\bar{3},1)]_A$ & $210$ & $(\bar{6},1)$ & $-4$ \\[3pt]

$[(\bar{3},1) \times (\bar{6},3)]_A$ & $210$ & $(\bar{6},3)$ & $-\frac{7}{3}$ \\[3pt]

$[(\bar{3},1) \times (\bar{6},3)]_S$ & $105$ & $(3,3)$ & $-\frac{7}{3}$ \\[3pt]

$(\bar{3},1) \times (6,1)$ & $\frac{1}{2}210 + \frac{\sqrt{3}}{2}105'$ & $(15+3,1)$ & $-1$ \\[3pt]

$(\bar{3},3) \times (6,3)$ & $\frac{\sqrt{3}}{2}210 - \frac{1}{2}105'$ & $(15+3,1)$ & $+\frac{1}{3}$ \\[3pt]

$(\bar{3},1) \times (\bar{3},3)$ & $\frac{1}{\sqrt{2}}210 - \frac{1}{\sqrt{2}}105'$ & $(15+3,3)$ & $-\frac{4}{3}$ \\[3pt]

$(\bar{3},3) \times (6,3)$ & $\frac{1}{\sqrt{2}}210 + \frac{1}{\sqrt{2}}105'$ & $(15+3,3)$ & $+\frac{1}{3}$ \\[3pt]

$(\bar{3},3) \times (6,3)$ & $ 105' $ & $(15+3,5)$ & $+\frac{1}{3}$ \\[3pt]

$[(\bar{3},3) \times (\bar{3},3)]_A$ & $105'$ & $(15'+\bar{6},5+1)$ & $+\frac{4}{3}$ \\[3pt]

$[(\bar{3},3) \times (6,1)]_A$       & $105'$ & $(15'+\bar{6},3)$ & $+\frac{5}{3}$ \\[3pt]

$[(\bar{3},3) \times (\bar{3},3)]_S$ & $\sqrt{\frac{2}{3}}105-\frac{1}{\sqrt{3}}\overline{15}$ & $(15,3)$ & $+\frac{4}{3}$ \\[3pt]

$[(\bar{3},3) \times (6,1)]_S$ & $\frac{1}{\sqrt{3}}105+\sqrt{\frac{2}{3}}\overline{15}$ & $(15,3)$ & $+\frac{5}{3}$ \\[6pt]
\hline
\end{tabular}

\caption{ $SU(6)_{CS}$ and $SU(3)_F \times SU(2)_S$ transformation properties  
and chromo-magnetic contribution to the mass splitting
for the two pair of diquarks in the $4q, L=1$ states}
\end{center}
\end{table}

From eqs.(1,2) and Tables 1 and 2 we derive the contribution to the  
pentaquark masses coming from the chromo-magnetic interaction. By including the 
contribution of the quark masses, the kinetic energy associated to the
angular momentum and the spin-orbit term, one gets:
\bea
m(p_1) &=& {\large \Sigma_{i=1,4}} \, m_{q_i} + m_{\bar{q}} +
\Delta m^1_{qq} + \Delta m^2_{qq} 
+ \nonumber \\
&&  C_{4q,\bar{q}} \left[ C_6(p)-C_6(t)-\frac{1}{3}C_2(p)+\frac{1}{3}
C_2(t)-\frac{4}{3} \right] + \nonumber \\
&& + K_1 + a \, \vec{L} \cdot \vec{S_q} 
\label{equation4}
\eea
where $\Delta m^1_{qq} + \Delta m^2_{qq}$ is the contribution of the 
chromo-magnetic interaction coming from the two {2q} clusters according to
eq.(1). 
A similar form holds for the positive parity states built with $4q$ in S-wave and
$\bar{q}$ in P-wave with respect to them:
\bea
m(p_2) &=&
{\large \Sigma_{i=1,4}} \, m_{q_i} + m_{\bar{q}}  + \nonumber \\
&& C_{qq} \left[ C_6(t)-\frac{1}{3} C_2(t)-\frac{26}{3} \right]
 + K_2 + b \,\vec{L} \cdot \vec{S_q} + c \,\vec{L} \cdot \vec{S_{\bar{q}}}
\label{equation5}
\eea
Finally for the negative parity states built with $4q$ and a $\bar{q}$ in S-wave
one has the mass formula:
\bea
m(s) &=&
 {\large \Sigma_{i=1,4}} \, m_{q_i} + m_{\bar{q}} +  C_{4q,\bar{q}}
 \left[ C_6(p)-C_6(t)-\frac{1}{3}C_2(p)+\frac{1}{3}
 C_2(t)-\frac{4}{3} \right] + \nonumber \\
&& C_{qq} \left[ C_6(t)-\frac{1}{3}
 C_2(t)-\frac{26}{3} \right] 
\label{equation6}
\eea
For the $Y=2$ states we are interested the $q$'s are $u$ or $d$ quarks and
$\bar{q}$ is $\bar{s}$ and the sum
\beq
\label{equation}
{\large \Sigma_{i=1,4}} \, m_{q_i} + m_{\bar{q}} =
m_N + m_K + 2\,C_{qq} + 4\,C_{4q,\bar{q}}
\label{equation7}
\eeq
By relating $C_{qq}$ and $C_{4q,\bar{q}}$ to the splittings in the $SU(6)_{FS}$
multiplets, one has:
\beq
C_{qq} = \frac{m_{\Delta} - m_N}{4} = 72,5 MeV 
\label{equation8}
\eeq
\beq
C_{4q,\bar{q}} = \frac{3}{16}\,(m_{K^*}-m_K) \simeq C_{qq}
\label{equation9}
\eeq
so that the r.h.s. of eq.(7) is
\beq
940+500+435=1875MeV
\label{equation10}
\eeq
$K_1$ will be fixed by identifying the lightest positive parity Y=2, I=0 state
with the $\Theta^+(1540)$, while the ratio $\frac{K_2}{K_1}$ is found by
considering that the kinetic energy term associated to the orbital motion
is inversely proportional to the reduced mass of the states in P-wave and
so we get:
\beq
\frac{K_2}{K_1}=\frac{4\,m_{u,d}+m_s}{4\,m_s}=\frac{7}{8}
\label{equation11}
\eeq
The last equality follows from taking the ratio $\frac{m_{u,d}}{m_s}
=\frac{5}{8}$.
We take $a=40MeV$, as in \cite{BS}, $b=20MeV$ and $c=50MeV$ since the 
giro-magnetic factors are inversely proportional to the masses.\\
With these values one finds for the $I=0,1,2$ states the spectrum described
in Fig.1-3 (4) for the positive (negative) parity states. In Fig 4 also the isospin
is reported above each line.

\begin{figure}
\begin{center}
\epsfig{file=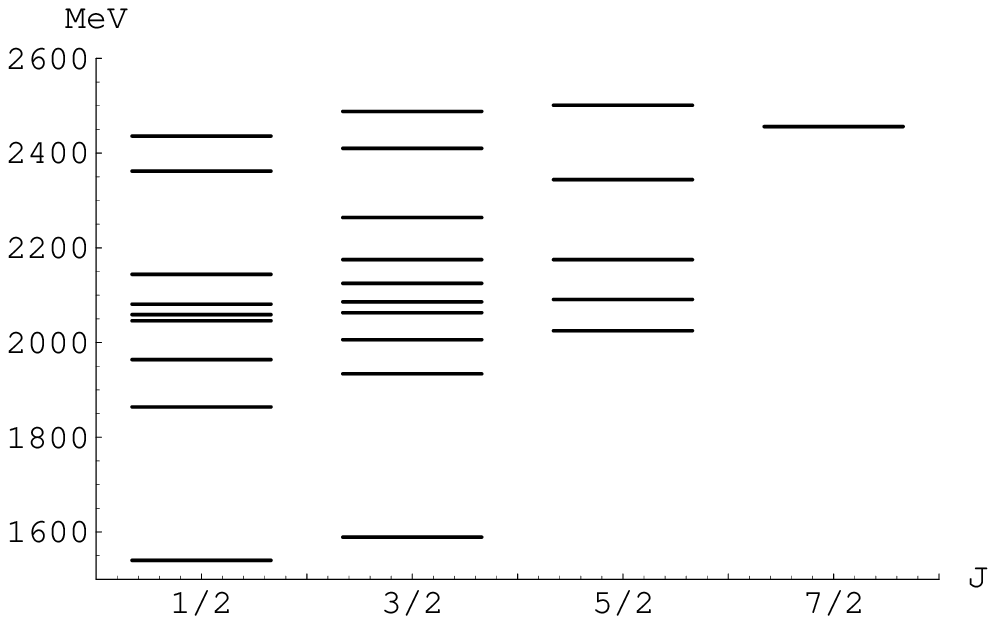,height=5cm}
\caption{The spectrum of the pentaquark with $I=0^+$}
\vspace{2cm}
\epsfig{file=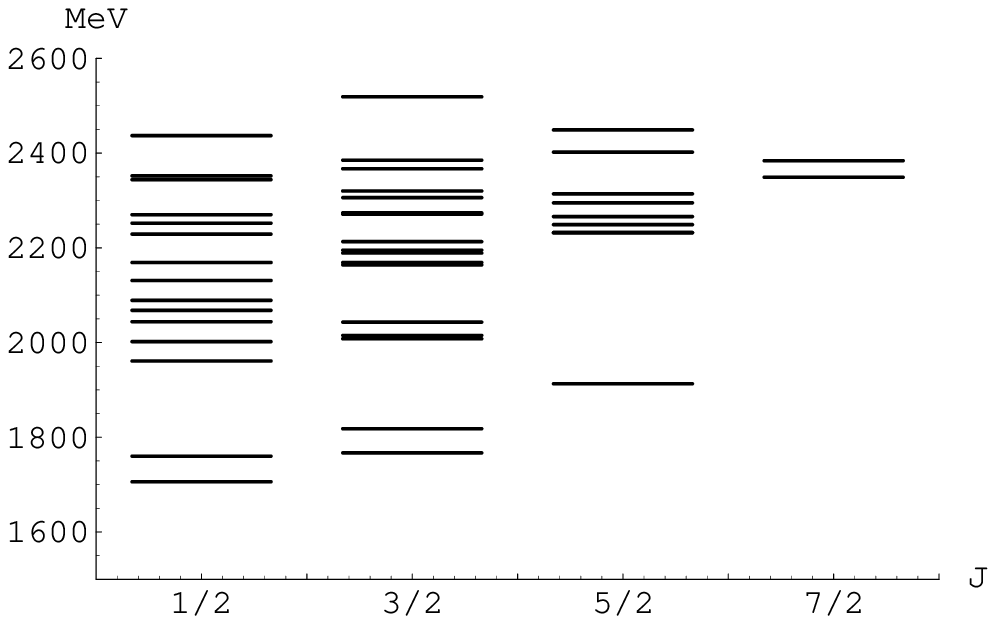,height=5cm}
\caption{The spectrum of the pentaquark with $I=1^+$}
\vspace{2cm}
\epsfig{file=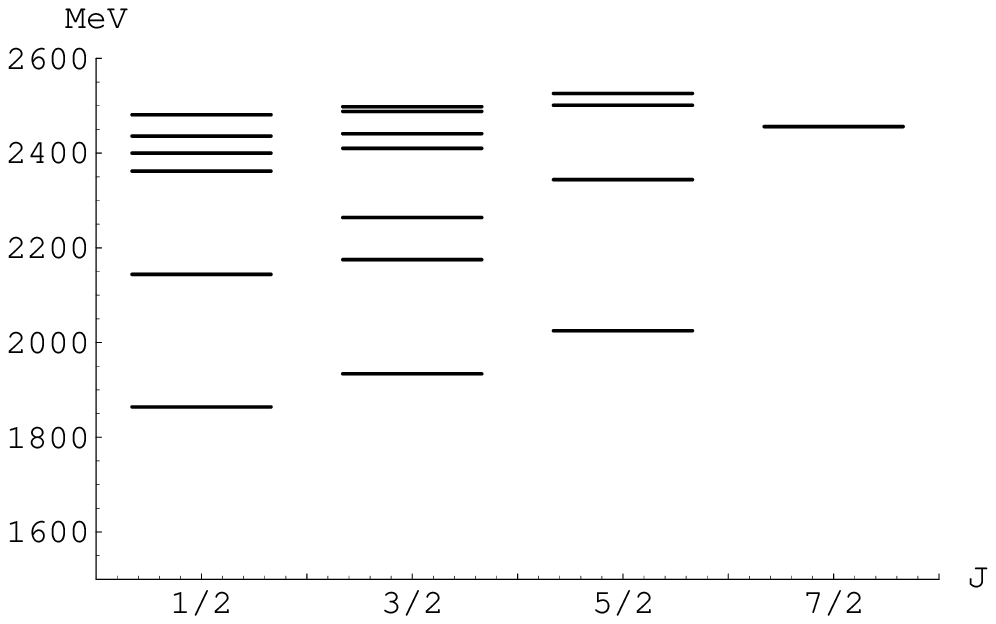,height=5cm}
\caption{The spectrum of the pentaquark with $I=2^+$}
\end{center}
\end{figure}

\begin{figure}
\begin{center}
\epsfig{file=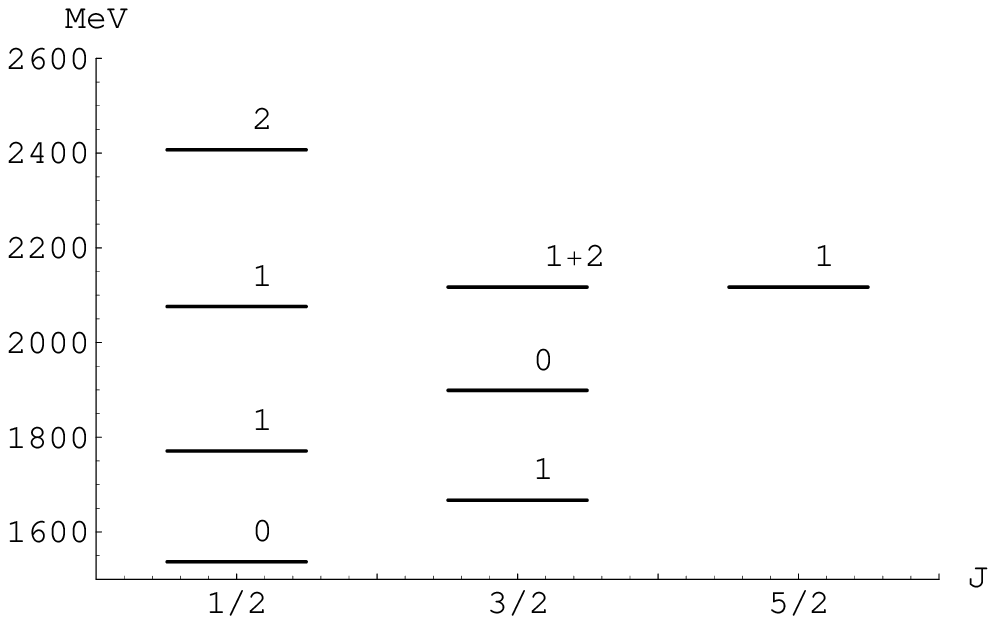,height=5cm}
\caption{The spectrum of the pentaquark with negative parity}
\end{center}
\end{figure}

The comparison  with experiment will be performed in the next section, after 
deducing the consequences of an approximate selection rule, one can reasonably 
assume for pentaquark decays.\\

\section{General properties of pentaquark decays}

The dynamics for pentaquark decays may be different from the case of
$\Delta$ and $\rho$ decays, where one has to create a $q \bar{q}$ pair,
with the $\bar{q}$ forming a meson with one of the initial quarks or with 
the initial quark.\\
As long as for pentaquarks, all the elementary fermions in the final state
are present in the initial one, which makes possible the hypothesis that
the decay is a consequence of the separation of its constituents. Also, 
at difference with what happens for the decay of the previously mentioned 
ordinary  hadrons, with the orbital angular momentum not conserved (changing
from 0 to 1) as well as the spin, for the decay of pentaquarks with $L=1$, to 
the initial orbital momentum of the $4q$'s in the initial state corresponds 
the relative angular momentum of the emitted meson with respect to the final 
baryon.
So $L$ and $S$ may be both conserved. One may even assume that, as in the 
hypothesis that the amplitude is proportional to the scalar product
of the initial and final wave-functions, also $SU(6)_{CS}$ and (or)
$SU(6)_{FS}$ \cite{GR} are conserved in pentaquark decays \cite{B}.
The consequences of $SU(6)_{CS}$ conservation for the decay of a 
pentaquark into a meson baryon final states are very restrictive
\cite{B}, since the pseudoscalar mesons are $SU(6)_{CS}$ singlets.
According to the $SU(6)_{CS}$ transformation properties of the
baryon of the 56 of $SU(6)_{FS}$, as a 70 the $(8,1/2)^+$ and as a 
20 the $(10,3/2)^+$, only the pentaquarks transforming as a 70 (20) 
of $SU(6)_{CS}$ may decay into a final state containing a pseudoscalar
meson and a $(8,1/2)^+$ ($(10,3/2)^+$) baryon. Let us therefore consider 
the $SU(6)_{CS}$ products:
\bea
210 \times \bar{6} &=& 1134 + 56 +70 
\label{equation12}
\eea
\bea
105 \times \bar{6} = 560 + 70
\label{equation13}
\eea
\bea
105' \times \bar{6}&=& 540 + 70 +20
\label{equation14}
\eea
\bea
\overline{15} \times \bar{6} = \overline{70} + 20
\label{equation15}
\eea
The combinations allowed to decay into a pseudoscalar meson 
and a $(8,1/2)^+$ (or a $(10,3/2)^+$) baryon, which transform as a 70 (20) of 
$SU(6)_{CS}$, are \cite{BS} for the $4q,L=1$ states:
\bea
&|70, &(1,S = 1/2), S_z = \frac{1}{2} > = \nonumber \\
& \frac{1}{\sqrt{3}} &  \{ \frac{1}{\sqrt{2}} \; | 210, (3, S = 1),
S_z = 1>_a \;
|\bar{6} , \, (\bar{3}, S= 1/2), S_z = - 1/2>^a \nonumber \\
& - \frac{1}{2} & |210, (3, S = 1), S_z = 0>_a \;
|\bar{6} , \, (\bar{3}, S = 1/2), S_z =  1/2>^a \nonumber \\
& + \frac{1}{2} & | 210, (3, S = 0) >_a \;
|\bar{6} , \, (\bar{3}, S = 1/2), S_z =  1/2>^a \} \nonumber \\
\label{equation16}
\eea
\newpage
\bea
&|70, & (1,S = 1/2), S_z = \frac{1}{2} > = \nonumber \\
& \frac{1}{\sqrt{3}} &  \{ \frac{1}{\sqrt{3}} \; | 105', (3, S = 1),
S_z = 1>_a \;
|\bar{6} , \, (\bar{3}, S= 1/2), S_z = - 1/2>^a \nonumber \\
& - \frac{1}{\sqrt{6}} & | 105', (3, S = 1), S_z = 0>_a \;
|\bar{6} , \, (\bar{3}, S = 1/2), S_z =  1/2>^a \nonumber \\
& + \frac{1}{\sqrt{2}} & | 105', (3, S = 0) >_a \;
|\bar{6} , \, (\bar{3}, S = 1/2), S_z =  1/2>^a \} \nonumber \\
\label{equation17}
\eea
\bea
&|20, & (1,S = 3/2), S_z = \frac{3}{2} > = \nonumber \\
&\frac{1}{\sqrt{3}} & \{ \frac{2}{\sqrt{7}} \; | 105', (3, S = 2), S_z = 2>_a \;
|\bar{6} , \, (\bar{3}, S= 1/2), S_z = - 1/2>^a \nonumber \\
& -\frac{1}{\sqrt{7}} & | 105', (3, S = 2), S_z = 1>_a \;
|\bar{6} , \, (\bar{3}, S = 1/2), S_z =  1/2>^a \nonumber \\
& +\sqrt{\frac{2}{7}} & | 105', (3, S=1), S_z = 1) >_a 
\;| \bar{6} , \, (\bar{3}, S = 1/2), S_z =  1/2>^a \} \nonumber \\
\label{equation18}
\eea
and the $S = 1/2~(3/2)$ state constructed by composing the $S=1$ of 
the $105 ~(\overline{15})$ with the $S=1/2$ of the $\bar{6}$;
$a = 1,2,3$ is a colour index to be saturated to get a colour singlet.
This motivates our attention to the $SU(6)_{CS}$ transformation 
properties of the pentaquarks. In fact the states, which transform as 
a $70(20)$, are allowed to decay into a $8,1/2^+(10,3/2^+)$ and a 
pseudoscalar meson, while the ones transforming as the
$\overline{70},540,560,1134$ representations are forbidden by the 
$SU(6)_{CS}$ selection rule to decay into those channels; we expect for 
them a three-body decay induced by the  $SU(6)_{CS}$  violating creation 
of a $q \bar{q}$ pair. In general the pentaquarks, which behave as a $70$ 
or a $20$ of $SU(6)_{CS}$, are lighter, since these representations have
lower $SU(6)_{CS}$ Casimirs.\\
In Table 3 we write for the positive parity states built with $4q, L=1$ and a 
$\bar{q}$ in S-wave the mass ( in $MeV$ ) of the $I=0$ states, transforming as 
a $70$, of the $I=1$, transforming as a $70$ or a $20$ and of the $I=2$ as a $20$, 
which are the only ones allowed by $SU(2)_I$ and $SU(6)_{CS}$ to decay into 
a meson and a baryon of the $56$ of $SU(6)_{FS}$.
Small mixing of states with the same $I, J$ quantum numbers are present.

\begin{table}[!h]
\begin{center}
\begin{tabular}{|c|c|c|c|}
\hline
& & & \\[-9pt]
$SU(6)_{CS} \times I$   &   $J=1/2$   &   $J=3/2$  &   $J=5/2$ \\
& & & \\[-9pt]
\hline
& & & \\[-9pt]
$(70,0)$              &   $1540$    &   $1589$   &   $$      \\[3pt]
$(70,1)$              &   $1706$    &   $1767$   &   $$      \\[3pt]
$(20,1)$              &   $1756$    &   $1818$   &   $1913$  \\[3pt]
$(20,2)$              &   $1864$    &   $1934$   &   $2025$  \\[3pt]
\hline
\end{tabular}
\caption{Positive parity states allowed to decay into a pseudoscalar meson
and a baryon of the $56$ of $SU(6)_{FS}$.}
\end{center}
\end{table}

As shown in \cite{CCKN2} also the states, which are allowed to have
two-bodies decays by the  $SU(6)_{CS}$ selection rule, may have a narrow 
width for a small overlap of the wave functions of the initial and the
final states.\\  
All the $I=2$ states are degenerate with $I=0$ states; as stressed by
Rossi and Veneziano \cite{RV}, the $I_3=0$ mass eigenstates have not 
definite isospin: in our case the diagonal states are the ones with
$(uu)(dd)$ or $(ud)(ud)$ clusters.\\ 
By considering ($4q$,$L=0$), we can build positive (negative) parity states 
by combining them with a $\bar{q}$ with $L=1(0)$ with respect to them. 
Within the approximation first suggested in \cite{HS} of requiring that
the $\bar{q}$ should form a meson only with a $q$ in the same cluster,
one expects a narrow width for the $J=1/2^+$ state built with 
($uudd,I=0,S=1,L=0$) and a $\bar{s}$ in P-wave respect to them.\\ 
In Table 4 we write the $SU(6)_{CS}$ transformation properties and the masses of
the negative parity states. 
\begin{table}[!h]
\begin{center}
\begin{tabular}{|c|c|c|c|}
\hline
& & & \\[-9pt]
$I$   & $SU(6)_{CS}$  $J=1/2$  & $SU(6)_{CS}$ $J=3/2$   & $SU(6)_{CS}$ $J=5/2$ \\[3pt]
\hline
& & & \\[-9pt]
$0$   & $70$          $1537$   & $560$        $1899$    & $$           $$      \\[3pt]
\hline
& & & \\[-9pt]
$1$   & $70$          $1537$   & $70$         $1537$    & $540$        $2117$  \\[3pt]
$ $   & $540$         $2076$   & $540$        $2117$    & $$           $$      \\[3pt]
\hline
& & & \\[-9pt]
$2$   & $\overline{70}$  $2407$  & $20$   $2117$    & $$     $$  \\[3pt]
\hline
\end{tabular}
\caption{Negative parity states.}
\end{center}
\end{table}
   
We expect that the states with $S=1/2$, which transform as the $70$ of 
$SU(6)_{CS}$ and are allowed to decay into the final state
consisting of a  pseudoscalar meson and a $1/2^+$  octet state, have a very 
large  width as the $SU(6)_{CS}$ singlet  \cite{J} $f^0(680)$ meson, 
which makes very difficult to identify them.\\
Instead the D-wave resonance, for which the orbital angolar momentum
changes by two units (from 0 to 2), are expected to have smaller
widths, which made possible their identification. In fact the only
negative parity $Y=2$ states, for which evidence has been found, are 
a $D03$ and a $D15$, both with a mass near to our prediction.\\
The lightest state has $I=0$ and $J^P=\frac{1}{2}^+$ and its
identification with $\Theta^+(1540)$ implies a value $K_1=308$ which is
a rather reasonable value. This state is almost the same considered by
previous authors \cite{JM} \cite{D}, where a variational approach has been 
developed to reach a higher precision for the determination of the spectrum. \\
Apart the fact of predicting a $I=0$ state as the lowest one with $Y=2$
the spectrum resulting from eq.(5-7) and Table 2 is in good agreement with
the evidence found for the $Y=2$ exotic states. In fact
the two $I=1$ positive parity states, $P_{11}(1720)$ and $P_{13}(1780)$
found in \cite{AR} are predicted properly . As long as negative parity 
states, one predicts a $D_{03}(1899)$ near to the $D_{03}(1865)$ found in 
\cite{PDG} and a $D_{15}(2117)$ to be compared with the evidence for a 
$D_{05}(2074)$ \cite{HARW} and (2150) \cite{PDG}.\\
To every $Y=2$ state, independently of its isospin, will correspond
a $Y=-1,I=\frac{3}{2}$ state. By neglecting the $SU(3)_F$ violation 
in the chromo-magnetic interaction, the spin-orbit term and the kinetic
energy, one expects a spectrum for these states given by the sum of
the spectra of the $Y=2, I=0,1,2$ states translated by the difference
of the constituent masses of the strange and light quarks, $175 MeV$.\\
So the $1862$ MeV $\Xi^{\pm}$ state seen by NA49 \cite{NA} could be 
identified with a member of the $27$ of $SU(3)_F$ representation, 
which contains the $P^{11}1720$.\\ 
We may conclude that the chromo-magnetic interaction, successful in reproducing
the mass splitting within the ordinary $SU(6)_{FS}$ hadron multiplets 
\cite{DGG}, seems rather promising to describe the spectrum of pentaquark
states.\\ 
The symmetry with respect to $SU(6)_{FS}$ would also have important
consequences. In fact by eqs.(12,15) and the tensor products:
\begin{equation}
56 \times 35 = 1134 + 700 + 70 + 56
\label{equation20}
\end{equation}
we reach the conclusion that the pentaquarks transforming as the
exotic $SU(3)_F$ representations cannot decay into the final
state consisting of a pseudoscalar or a vector meson  and a
baryon of the octet $1/2©^+$ or of the decuplet $3/2^+$, if
their $4q$'s transform as the $105 + 105'$ of $SU(6)_{FS}$
\cite{SR}, have their couplings to these states proportional
with ratios dictated by $SU(6)_{FS}$ symmetry if their $4q$'s
transform as the $126$ or $210$.\\

\section*{Acknowledgements}

Its a pleasure for one of us (F.B.) to acknowledge very instructive
discussions with Prof. H. H\"{o}gaasen and P. Sorba at LAPP and the nice 
hospitality received there, where this work began.

\end{document}